# DRU-Mixer: An Explainable Dilated Recurrent Unit-Based Mixer Model with Focal Loss for EEG-Based Interictal Epileptiform Discharges Detection

Aditya P Kamath[1*], Xiaohui Tao[1], Taotao Cai[2], Prabal Datta Barua[1], Sreeni Chadalavada[3], Abdulkadir Sengur[4]

[1]School of Mathematics, Physics and Computing, University of Southern Queensland, Springfield Central, QLD 4300, Australia
[2]School of Mathematics, Physics, and Computing, University of Southern Queensland, Toowoomba, Queensland, Australia
[3]School of Engineering, University of Southern Queensland, Springfield, Australia
[4]Electrical-Electronics Engineering Department, Technology Faculty, Firat University, Elazig, Turkey

*Corresponding author: Aditya P Kamath (e-mail: adityaprabhakara.kamath@unisq.edu.au).

## Abstract

Interictal epileptiform discharges (IED) have been identified as a core electroencephalogram (EEG) biomarker in diagnosing epilepsy patients. However, automatic detection and classification of IED require significant effort by human experts and can be prone to inter-rater variability. This paper presents Dilated Recurrent Units (DRU)-Mixer, a deep learning network to automate the process of identifying and classifying IED based on EEG recordings. The proposed architecture comprises a convolutional front end for extracting local features, dilated recurrent units modeling the temporal relationship between EEG signals, and mixer modules for token- and channel-level representation learning. DRU-Mixer was trained to perform (i) binary classification of IED/Non-IED segments and (ii) multiclass classification of regional IED types. For binary classification, DRU-Mixer showed 95.19% accuracy, 72.76% precision, 82.31% recall, 96.61% specificity, and 77.24% F1-score. For multiclass classification, the proposed network achieved strong sensitivity between 74.8% and 89.2%, as well as precision between 70.5% and 86.3%. Sensitivity-oriented analysis of the results shows that, for 70% sensitivity, DRU-Mixer provided 83.8% precision and 98.5% specificity outperforming VGGEEG and VGGSleep models and being on par with vEpiNet model. Interpretability of the model's results was estimated through occlusion mapping technique, which was used to highlight time intervals and EEG channels providing maximal contribution to the prediction. Maps generated by the technique show specific patterns for IED and Non-IED indicating that the model utilizes structured channel-time features corresponding to the presence of epileptiform activity.

# 1. Introduction

IED are intermittent abnormal waveforms detected using EEG and represent the hyper-excitability of neurons, which are primarily linked with the condition known as epilepsy. Epilepsy is a brain disease that affects more than 50 million people globally and involves unprovoked attacks of seizures, which could have a profound effect on the quality of life. EEG is a non-invasive test that allows recording of electrical activity of the brain and continues to be considered the gold standard in identifying IED, playing an important role in diagnosis, classification, and treatment of epilepsy. IED need to be detected accurately due to their scarcity and the high risk of interference with artifacts, resulting in prolonged manual analysis that can be subjective. Epileptiform discharges constitute a very small fraction of the total recordings, which poses a challenge in class imbalance during the EEG analysis, this imbalance has shown a significant effect on the model sensitivity and generalization [29]. Significant progress has been made in deep learning and signal processing, greatly enhancing the accuracy of automated detection of IED.

In view of the complicated nature and volume of EEG data, it is apparent that artificial intelligence (AI) is the critical tool necessary to help physicians interpret IED to detect epilepsy [6]. Human interpretation of long-term or high-density EEG recordings is laborious, and it can be subject to human error. In contrary, AI can process large volumes of data promptly and uniformly [7]. Deep learning algorithms, such as convolutional neural networks (CNN) and transformers, have shown significant performance in detecting epileptiform discharges that can be difficult to identify during conventional examination [8]. In addition to reducing the burden of making a diagnosis, AI systems allow the decision-making process in cases when the situation is not clear, which increases efficiency and reduces the margin of error in clinical practice. The use of an AI-assisted IED detection system helps in shifting clinicians' attention towards treatment options and patient care [9].

Several AI-based methods for IDEs-based epilepsy detection have been reported in the literature. For example, Lourenço et al. [10] investigated the impact of various EEG data types on the performance of deep learning for IED detection. They utilized a VGG-C CNN model with some adjustments that was trained on three datasets: routine EEGs, ambulatory EEGs, and a combined dataset comprising data from 308 patients recorded at Medisch Spectrum Twente. The model on normal EEGs had 84% sensitivity at 99% specificity for normal tests, but a poor 53% sensitivity on ambulatory data. The multi-data- and ambulatory-trained models performed better on ambulatory EEGs (60% and 79% sensitivity, respectively, at 99% specificity) but showed sensitivity below 60% on normal EEGs. The main limitation is that model performance is highly environment-dependent, resulting in weak cross-context generalization. Dissanayake et al. [11] proposed two patient-independent deep learning models to predict seizures from scalp EEG recordings of the CHB-MIT database, aiming to detect the pre-ictal brain state within a one-hour time window. The authors experimented with a common CNN and a Siamese model, utilizing a novel learning strategy to address severe inter-subject variability. Accuracy rates of 88.81% and 91.54% were achieved, representing improvements compared to the existing method. In addition, the Siamese model showed itself to be trainable for individual patient predictions, with an accuracy rate of 97%. The authors have used model interpretation to explain how the models learn. The main issue is that while the findings are promising, actual implementation might face challenges due to potential differences in unknown data and acquisition conditions. Thomas et al. [12] developed an EEG classification system aimed at identifying the existence of epileptic recordings by detecting IED. This system uses a CNN for classification on the level of waveforms and a support vector machine (SVM) for classification on the level of EEGs. The data used were taken from 156 EEGs from Massachusetts General Hospital and were labeled for 18,164 spikes. The accuracy rate of the system in classification of EEGs with and without IED reached 83.86%. The main limitation is relatively modest accuracy, and the approach is heavily dependent on the quality of spike annotations, with class imbalance being a long-standing issue, even with balanced training methods. Ouichkan et al. [13] proposed five deep learning architectures, four different models, and a ResNet50 with transfer learning for seizure prediction on intracranial EEG data from the American Epilepsy Society Seizure Prediction Challenge. The two CNN models had the best performance with 95% accuracy. The limitation is that, although the dataset achieved high accuracy, the models were trained on a relatively small number of subjects, which raises questions about their generalizability to different populations and clinical conditions. Sun et al. [14] suggested U-IEDNet, a deep learning architecture that combines convolutional layers, bidirectional GRUs, and transformer networks to classify and segment IED in multichannel EEG while preserving spatial and inter-channel relationships. It was trained and tested on the publicly available TUEV database (370 patients) and also on a smaller annotated dataset (43 patients). On TUEV, U-IEDNet achieved a recall of 0.916, a precision of 0.911, and an F1-score of 0.912, with acceptable segmentation metrics (F1-score of 0.909). On their in-house dataset, the classification F1-score was 0.903. Although interpretability was improved by attention weight analysis, model complexity and computational demands may limit its deployment in low-resource clinical setups. Lin et al. [15] proposed an open-access EEG dataset containing comprehensive spatial distribution information of IED, as well as sleep–wake labels. The data were taken from 84 patients, with each patient having 20 minutes of continuous EEG, for a total of 28 hours and offering 2,516 IED epochs and 22,933 non-IED epochs. The VGG18 detection algorithm with spatial and consciousness-state features yielded better performance. The same dataset was used in order to conduct clinical tests and prove its applicability in the context of the objective evaluation of detection algorithms. One drawback associated with using this dataset is the presence of artifacts related to environment and movement, which may lead to higher false-positive rates. Chung et al. [16] proposed using deep learning techniques for focal IED detection and classification in scalp EEG recordings from the frontal, temporal, and occipital brain areas. Analysis of 1.5 seconds of EEG samples was conducted using one- and two-dimensional convolutional neural networks on a sample consisting of 38 patients with focal epilepsy and 232 controls. Binary classifiers showed accuracy ranging from 79.3% to 86.4% (frontal), 93.3% to 94.2% (temporal), and 95.5% to 97.2% (occipital), while 3- and 4-class classifiers demonstrated accuracy of 87.0% to 88.7% and 74.6% to 74.9%, respectively. A major drawback was the lower accuracy for detection of frontal IED using multiclass classification algorithms. Zhou et al. [17] have created an extensive EEG dataset which is meant to provide the opportunity for training and cross-dataset validation of artificial intelligence-based epilepsy diagnostics models. The dataset includes thousands of labeled IED and non-IED segments with various types of patients along with additional information such as the location of electrodes and patient demographic data. Performance baseline for standard CNN model looked rather promising although limitations in accuracy due to imbalance and overfitting risks exist. Bagheri et al. [18] have proposed a fast machine learning technique for IED detection using scalp EEG by filtering out most of the background waves prior to running expensive computation-intensive classifiers. The technique uses an ensemble of simple classifiers based on feature selection in order to exclude 78% of background waves while retaining 97% of IED in the sample including 156 EEGs from Massachusetts General Hospital. The use of such preprocessing resulted in an increase in speed for subsequent classification with SVM, k-NN, and random forests ranging from 1.8 to 4.7 times. While the process is fast, it requires accurate annotations initially, making it less generalizable to new datasets with different characteristics of EEG data. Lourenço et al. [19] studied the effect of augmentations, such as temporal shifts and EEG montages, on the performance of CNN in detecting IED. Applying the technique to EEG from 166 patients with different types of epilepsy and controls, the researchers found out that augmentations helped increase sensitivity from 63% to 96% at 99% specificity, with false positives decreasing from 2.11 to 0.73 detections per minute. It helped in reducing the requirement for a large amount of annotated data by experts, but the performance depends on good initial annotations and may be dataset dependent. In their work on 2021, Westover et al. explored the successful use of clinical EEG records in deep learning applications for epilepsy and considered ways to train CNN using small and annotated datasets. Applying the same augmentation approaches used in [19], they showed that the approach reduces the need for labor-intensive labeling by experts and preserves good detection performance. Their CNN network, based on the VGG architecture, showed promising performance improvement after augmentation, but, like other research, the robustness of this network in different real-world conditions remains to be proved. Stamoulis et al. [21] proposed a deep-learning-based method to automatically detect impulsive events of interest by training CNN with extensive annotated datasets collected from various centers. This method uses raw EEG waveform data without the use of any manual features and shows superior sensitivity and specificity compared to traditional methods. The performance of this method is shown to be competitive on various heterogeneous datasets, which implies generalization capability; nevertheless, the differences in EEG recording techniques and annotation criteria across different sites can pose a limitation to universality.

This paper introduces an EEG-based lightweight IED detector comprising a convolutional front-end, a dilated bidirectional recurrent unit, and two Mixer blocks (DRU-Mixer) that simultaneously learn long-term temporal patterns and cross-channel relationships. The end-to-end pipeline encompasses data pre-processing and classification. Multi-channel EEG windows are band-pass filtered to mitigate drift and high-frequency noise, and each window is normalized per channel to standardize amplitude variance. A two-step one-dimensional convolutional stem maps raw traces onto feature maps, which are processed further as a sequence and fed into a multi-layer dilated bidirectional recurrent unit. Temporal dilation expands the effective receptive field without increasing parameter count. In addition to the recurrent states, two Mixer blocks perform token mixing across time and channel mixing across features to produce an expressive sequence representation. Dropout regularization and a fully connected layer generate binary logits for IED vs. Non-IED classification and multi-class classification. Class imbalance is addressed through weighted sampling of the training data combined with focal loss, where the weighting factor reflects the observed class ratio. The model is trained with the Adam optimizer, gradient clipping, and a learning rate scheduler that reduces the rate when progress plateaus. Evaluation is performed on a stratified split of the data. At test time, Monte Carlo dropout with multiple stochastic forward passes provides uncertainty-informed mean probabilities for reporting accuracy, precision, recall, F1-score, sensitivity, specificity, and the confusion matrix. Additionally, a sensitivity-targeted analysis sweeps decision thresholds to estimate operating points across a range of desired sensitivities, reporting the corresponding precision, specificity, and threshold for clinically tuned deployment.

The main contributions of this paper are:

1) An end-to-end differentiable DRU-Mixer architecture incorporating convolutional feature extraction, dilated bidirectional recurrent temporal modeling, and token/channel mixing for multichannel EEG.

2) A realistic training approach that addresses class imbalance through a weighted sampling approach using focal loss, made stable through the use of gradient clipping and an adaptive learning rate.

3) Uncertainty-based inference using Monte Carlo Dropout along with a thresholding strategy that highlights the critical operating point for clinical needs.

## 2. Materials and Methods

This research study is concerned with the design and evaluation of the DRU-Mixer model for automatic detection of IED from EEG signals. In the proposed method, convolutional layers have been used for extracting features, DRU for modeling long-range dependencies, and mixer layers for better token and channel interaction. Raw EEG signals undergo preprocessing procedures, such as bandpass filtering and standardization, to improve noise reduction and normalize activity across channels. Data are organized into Non-IED and IED classes, with stratified train-test splits, and class imbalance is corrected with a weighted sampling approach. The model is trained using Focal Loss to prioritize hard-to-classify examples and optimized using Adam, along with a learning rate scheduler. Fig. 1 shows the schematic illustration of the proposed method.

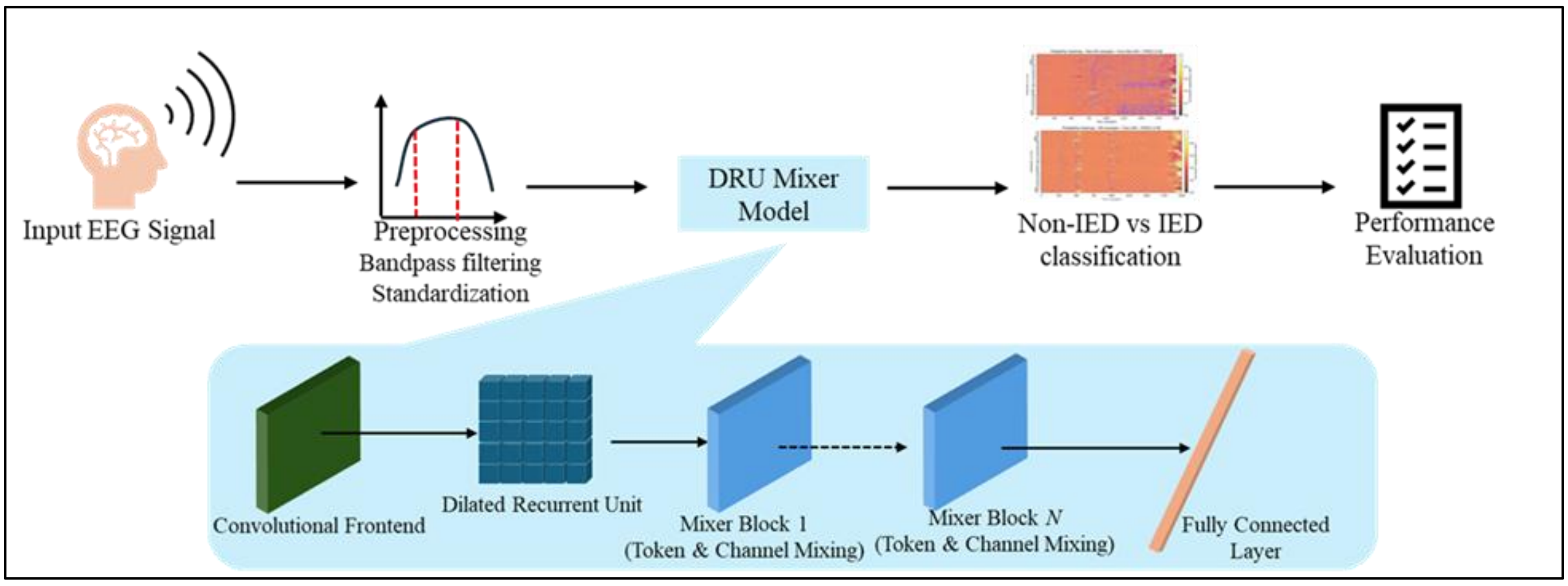

**Figure 1**: Illustration of the proposed method.

## 2.1 Dilated Recurrent Unit

Let a multichannel sequence be $x_{\{1:T\}}, x_t \in \mathbb{R}^C$. A dilated recurrent unit with dilation factor r replaces the nearest-neighbor temporal dependency with a stride-r dependency [22]. The recurrent state at time t depends on the state r steps back.

$$h_t = h_{t-r}$$

Using GRU-style gating with dilation [23]:

$$z_t = \sigma(W_z\, x_t + U_z\, h_{t-r} + b_z)$$

$$r_t = \sigma(W_r\, x_t + U_r\, h_{t-r} + b_r)$$

$$\tilde{h_t} = tanh(W_h\, x_t + U_h\, (r_t \odot h_{t-r}) + b_h)$$

$$h_t = (1 - z_t) \odot h_{t-r} + z_t \odot \tilde{h_t}$$

with $h_t = 0$ for $t \le 0$ (or a learned initial state). When $r = 1$, the unit reduces to a standard Gated Recurrent Unit (GRU). The effective receptive field of $h_t$ includes inputs $\{x_t, x_{t-r}, x_{t-2r} \ldots\}$, so the model covers a temporal span of rL steps with only L recurrent transitions. This yields longer-range context with fewer sequential dependencies, reducing gradient attenuation and wall-clock cost.

A practically equivalent view is time subsampling: define $\hat{x}_k = x_{1+r(k-1)}$ for $k = 1, \ldots, T' = \lceil T/r \rceil$, and apply a standard GRU over $\hat{x}_{\{1:T'\}}$. Both realizations perform $O(T/r)$ recurrent updates and maintain the same parameter count as the base GRU. In bidirectional form, forward and backward dilated states are defined in Eq.1;

$$\overrightarrow{h_t} = DRU_r(x_t, \overrightarrow{h}_{t-r}) \qquad (1)$$

$$\overleftarrow{h_t} = DRU_r(x_t, \overleftarrow{h}_{t+r})$$

These states may be concatenated as $h_t^* = [\overrightarrow{h_t}; \overleftarrow{h_t}] \in \mathbb{R}^D$. A sequence-level descriptor may be taken as the final valid index after dilation, e.g., $s = h^*_{T'}$, on the subsampled timeline of length T′.

## 2.2 Dilated Recurrent Unit Mixer

Let $H \in \mathbb{R}^{T' \times D}$ be the output sequence of a bidirectional dilated recurrent encoder (e.g., concatenated forward–backward DRU states), where $T' = \lceil T/r \rceil$. A DRU–Mixer block augments recurrent modeling with two residual mixing operations that act along complementary axes [24]:

1) Channel mixing (per time step) applies a position-wise MLP to each row of H [25]:

$$\hat{H} = LN(H)$$

$$CM(H)_t = \phi(\hat{H}_t W_{c1} + b_{c1})W_{c2} + b_{c2}, \quad t = 1, \dots, T' \tag{2}$$

$$U = H + CM(H)$$

Where $\hat{H}_t \in \mathbb{R}^D$ is the t-th row, $W_{c1} \in \mathbb{R}^{D \times d_c}$, $W_{c2} \in \mathbb{R}^{d_c \times D}$, $\phi$ is a pointwise nonlinearity (e.g., GELU), and LN denotes layer normalization along the feature dimension. Channel mixing reweights and mixes features within each time frame, conditioned on local context via LN.

2) Token mixing (per feature channel) mixes information across time for each column of U. Writing, $U^T \in \mathbb{R}^{D \times T'}$ and operating columnwise in the temporal axis:

$$\tilde{U} = LN(U)$$

$$TM(H)_{:d} = \phi(\tilde{U}_{:d} A_{t1} + a_{t1})A_{t2} + a_{t2}, \quad d = 1, \dots, D \tag{3}$$

$$V = U + TM(u)$$

with $\widetilde{U_{:d}} \in \mathbb{R}^{T'}$, $A_{t1} \in \mathbb{R}^{T' \times d_t}$, $A_{t2} \in \mathbb{R}^{d_t \times T'}$. This step enables non-recurrent, all-to-all temporal interactions at each feature channel, complementing the DRU's sequential pathway.

Stacking $L$ such Mixer blocks yields:

$$H^{(0)} = H$$

$$H^{(\ell+1)} = Mixer(H^{(\ell)}), \quad \ell = 0, \dots, L-1 \tag{4}$$

preserving sequence length $T'$ and feature width $D$ through residual connections, which stabilize optimization and improve gradient flow. A classifier then consumes a sequence summary, for example, the last DRU–Mixer state;

$$y = W_o\, Dropout(H_{T'}^{(L)}) + b_o \in \mathbb{R}^K \tag{5}$$

where K is the number of classes. The division of labor is thus explicit: the dilated recurrent encoder provides long-range temporal sensitivity with $OT/r$ recurrent steps and gated memory, while the token mixer supplies non-recurrent global temporal aggregation and the channel mixer refines per-frame feature interactions. Together, they form a compact architecture that captures multiscale temporal structure without inflating parameter count or sequential depth.

## 2.3 Dataset

The dataset that was used for the analysis in this study was obtained from EEG data collected from patients at Seoul National University Bundang Hospital [16]. The dataset includes 38 focal epilepsy patients (15 females and 23 males with an average age of 20.1 ± 16.2) and 232 control subjects (132 females and 100 males with an average age of 12.9 ± 3.1). Controls were selected among people who had paroxysmal non-epileptic neurological symptoms like syncope, headaches, or vertigo but did not have any seizures or neurological disorders. Patients with focal epilepsy were further categorized into three groups according to the location of IED: 15 with frontal IED (three females and 12 males, mean age 11.4 ± 3.6 years), 13 with temporal IED (seven females and six males, mean age 35.3 ± 20.2 years), and 10 with occipital IED (five females and five males, mean age 13.2 ± 2.7 years).

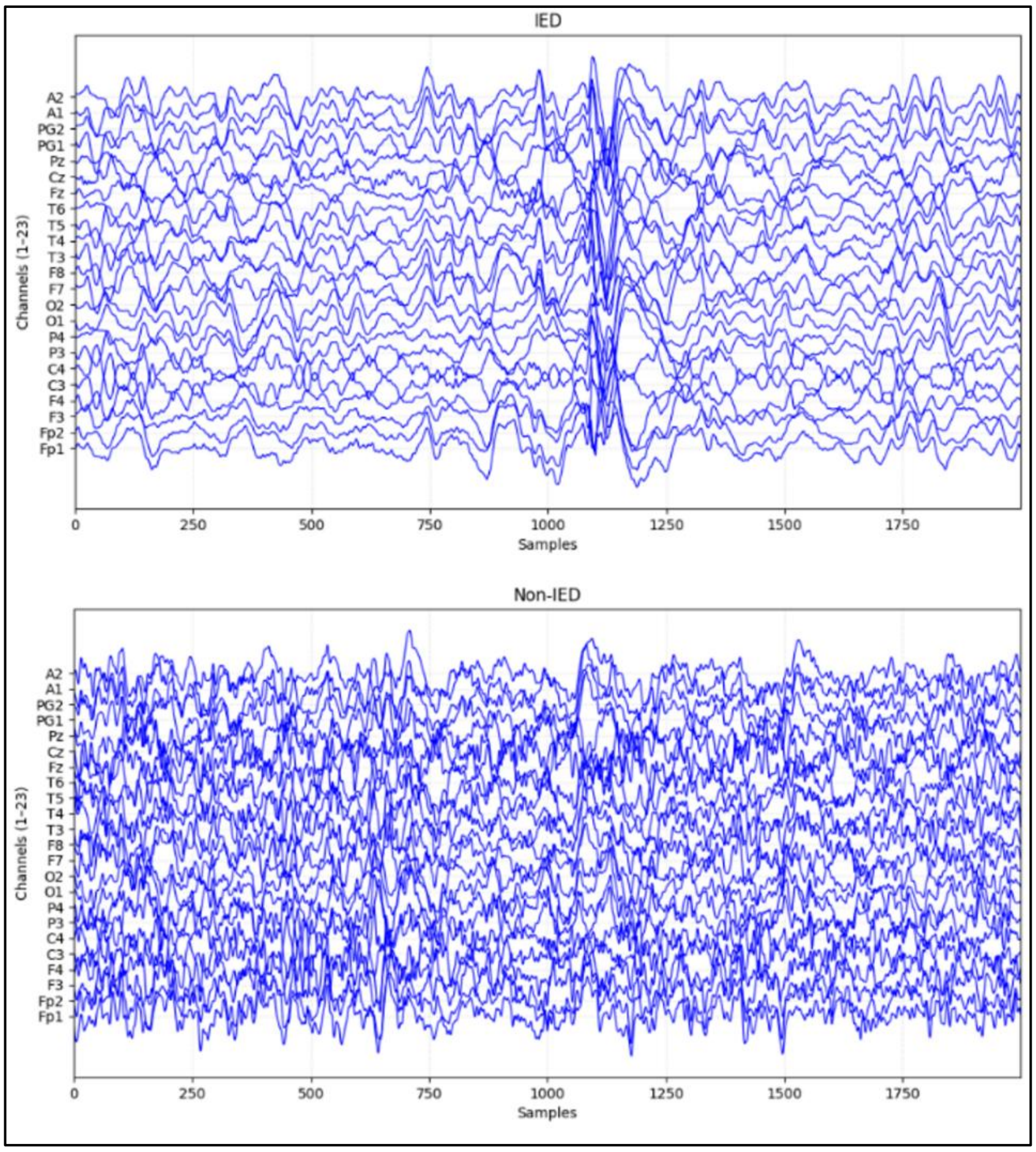

**Figure 2:** IED and Non-IED EEG Samples.

EEG recordings were obtained during both awake and resting states, as well as during sleep, using a 23-channel digital EEG system, following the international 10–20 electrode placement scheme. Each session lasted at least 30 minutes, with a sampling frequency of 200 Hz, a band-pass filter of 1–70 Hz, and a 60 Hz notch filter. Nineteen electrodes were used, and recordings were re-referenced to the average montage. When clinically necessary, pediatric patients received chloral hydrate as a sedative. A total of 4557 IEDs were annotated across the patient group, consisting of 2112 frontal, 1176 temporal, and 1269 occipital discharges.

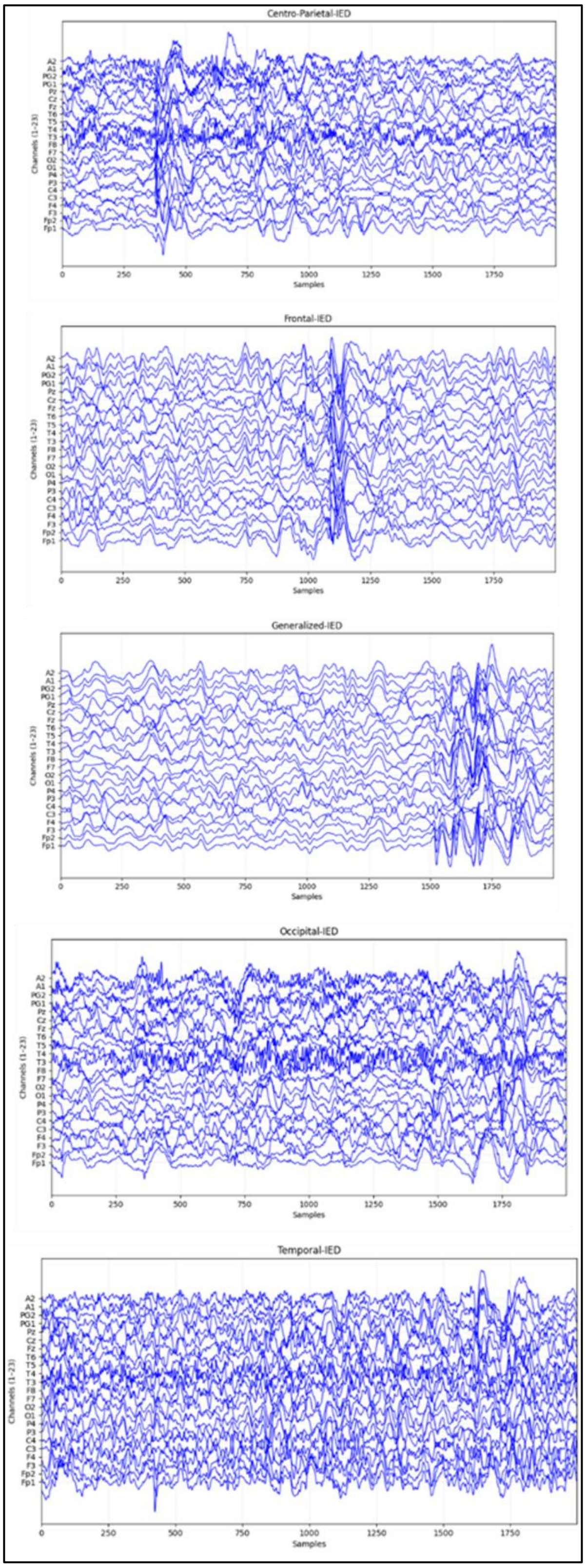

**Figure 3**: Multi-Class IED EEG Samples.

On average, there were 141 ± 129 IED per recording in the frontal group, 90 ± 99 in the temporal group, and 127 ± 68 in the occipital group. Two epileptologists reviewed all patient EEGs to identify IED, defining each from the onset of a spike, sharp wave, or spike-and-wave complex to the end of the discharge. This dataset is particularly useful for machine learning applications because it supports two different classification settings. In the first setting, EEG segments are organized as a binary classification problem to distinguish IED from non-IED segments. In the second setting, only IED segments are considered and classified according to their spatial distribution. In this study, the multi-class IED classification task included five spatial IED categories: frontal, temporal, occipital, generalized, and centro-parietal IED. Thus, the binary task focused on detecting the presence of IED activity, whereas the multi-class task focused on identifying the spatial subtype of IED events. Fig. 2 shows representative EEG samples for the IED and non-IED binary classification task, while Fig. 3 shows representative EEG recordings for the five IED classes used in the multi-class classification task.

# 3. Experimental work and results

The proposed DRU-Mixer architecture was implemented in Python 3.10 using PyTorch 2.2, a deep-learning framework. All experiments were conducted on a workstation equipped with an NVIDIA GeForce RTX 3090 GPU (24 GB VRAM), an AMD Ryzen 9 5950X CPU, and 128 GB RAM. The dataset was split into disjoint partitions using a segment-level stratified random split, with 80% of the EEG recordings used for training and 20% for independent testing, ensuring stratification to preserve the relative class balance between IED and non-IED events. Raw EEG matrices of shape $23 \times 2000$ samples were first zero-phase band-pass filtered between 1 and 40 Hz using a fourth-order Butterworth filter, then each channel was standardized by z-scoring. The resulting trials were transposed to a $T \times C$ format and stored as PyTorch tensors. The DRU-Mixer model begins with a pair of temporal convolutional layers (with kernel sizes of 7 and 5, and 64 feature maps each) followed by a DRU with a hidden size of 128 and a dilation factor of 2, which captures long range temporal dynamics. Two stacked Mixer blocks perform token- and channel-wise feature mixing, after which a dropout layer ($p = 0.5$) and a fully connected layer produce the class logits. Training was performed using mini-batches of 32 with weighted random sampling to counteract class imbalance, the Adam optimizer (learning rate 1e-4), and a focal-loss objective ($\alpha$ scaled to the class ratio, $\gamma = 2$. The network was trained for 70 epochs with gradient clipping at norm 1.0. During evaluation, Monte-Carlo dropout with 10 stochastic passes was employed to estimate predictive uncertainty. Performance on the test set was reported in terms of accuracy, precision, recall, F1-score, and specificity.

The DRU-Mixer model begins with a pair of temporal convolutional layers (with kernel sizes of 7 and 5, and 64 feature maps each) followed by a DRU with a hidden size of 128 and a dilation factor of 2, which captures long-range temporal dynamics. Two stacked Mixer blocks perform token- and channel-wise feature mixing, after which a dropout layer ($p = 0.5$) and a fully connected layer produce the class logits. Training was performed using mini-batches of 32 with weighted random sampling to counteract class imbalance, the Adam optimizer (learning rate 1e-4), and a focal-loss objective ($\alpha$ scaled to the class ratio, $\gamma = 2$). The network was trained for 70 epochs with gradient clipping at norm 1.0. During evaluation, Monte-Carlo dropout with 10 stochastic passes was employed to estimate predictive uncertainty. Performance on the test set was reported in terms of accuracy, precision, recall, F1-score, and specificity.

## 3.1 Non-IED vs IED Classification Task

For the binary classification task, IED and Non-IED segments were used as two separate classes. The Non-IED samples were obtained from the pre-organized Non-IED class of the dataset and represent EEG segments without annotated IED activity. No additional manual selection of Non-IED samples was performed in this study. All available Non-IED segments with the expected data shape were included and processed using the same preprocessing pipeline as the IED samples, including band-pass filtering and channel-wise standardization.

The classification of Non-IED vs. IED was performed, and the related performance evaluation metrics are presented in Table 1. As indicated in Table 1, the model achieved an accuracy of 95.19%, reflecting good reliability in discriminating between the two classes. 72.76% precision indicates the moderate rate of correctly identified IED samples among all estimated positives, and 82.31% recall indicates the ability of the model to detect most actual IED samples. The model's 96.61% specificity indicates its effectiveness in accurately detecting non-IED samples, thereby minimizing false positive detections. The F1-score of 77.24% provides an accurate and balanced estimate of precision and recall.

**Table 1**: Performance evaluation metrics for a binary classification task (Non-IED vs IED).

| | **Accuracy (%)** | **Precision (%)** | **Recall (%)** | **Specificity (%)** | **F1score (%)** |
|---|---|---|---|---|---|
| **DRU-Mixer** | 95.19 | 72.76 | 82.31 | 96.61 | 77.24 |

Fig. 4 shows the confusion matrix that the DRU-Mixer has revealed. As is evident from Fig. 4, 4419 non-IED samples and 414 IED samples were correctly classified, while 155 non-IED and 89 IED samples were incorrectly classified.

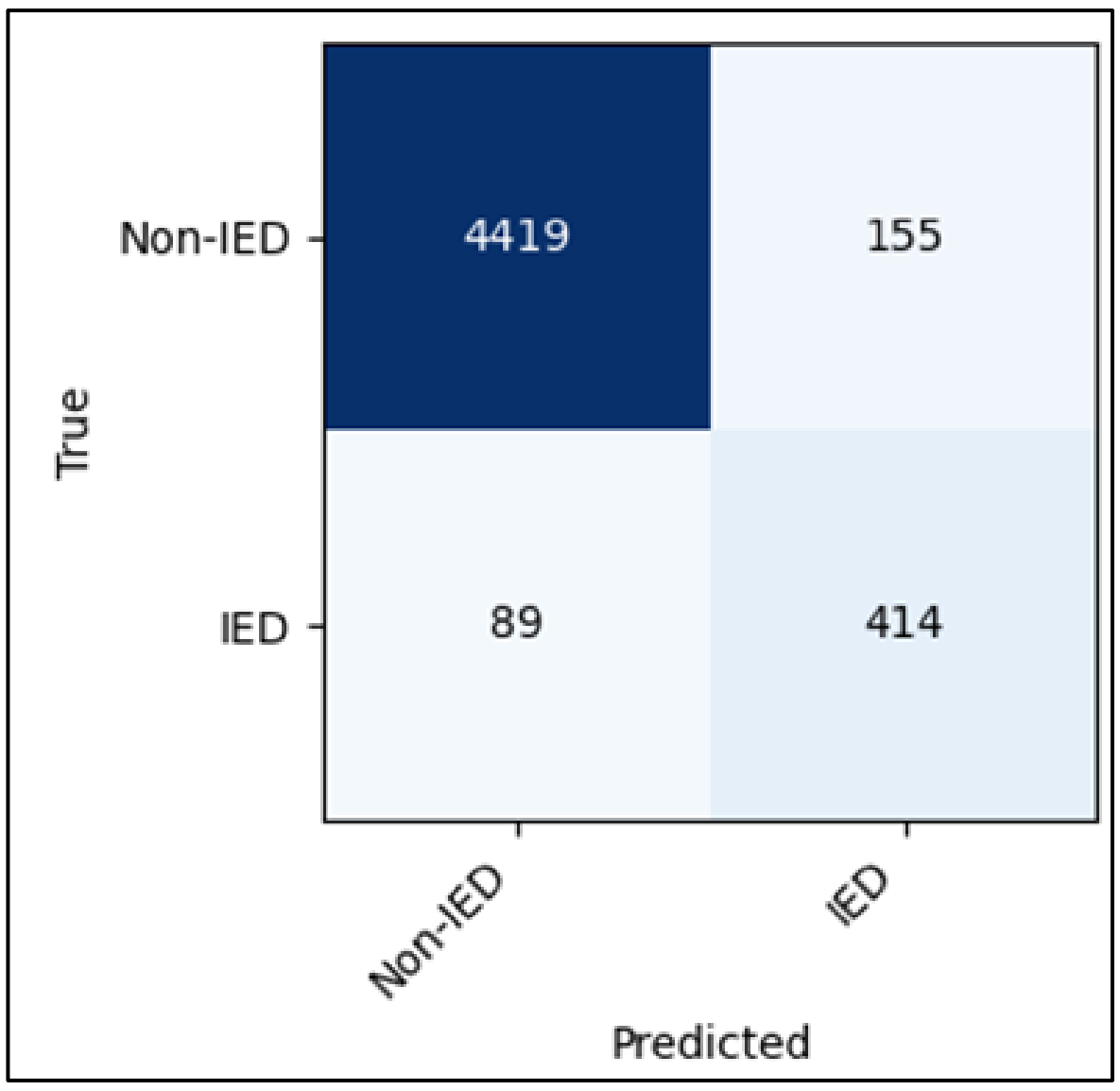

**Figure 4**: Obtained confusion for Non-IED vs IED classification task.

Table 2 presents the sensitivity, specificity, and precision of the proposed DRU-Mixer model at different levels of sensitivity. For lower sensitivity levels (50–70%), the precision (93.9%–83.8%) and specificity (99.7%–98.5%) of the model are high, indicating good detection of IED events and minimal false positives. As the level of sensitivity increases beyond 75%, precision and specificity gradually fall, with precision reaching 49.3% and specificity reaching 89.5% at a 95% sensitivity level. This is the compromise that reveals the model's tendency to generate more false positives to cover a greater number of true IED cases.

**Table 2:** Sensitivity, specificity, and precision of the proposed DRU-Mixer model.

| Sensitivity (%) | Precision (%) | Specificity (%) |
|---|---|---|
| 50 | 93.9 | 99.7 |
| 55 | 91.7 | 99.5 |
| 60 | 89.3 | 99.2 |
| 65 | 87.0 | 98.9 |
| 70 | 83.8 | 98.5 |
| 75 | 81.7 | 98.2 |
| 80 | 76.9 | 97.4 |
| 85 | 66.6 | 95.3 |
| 90 | 54.8 | 91.8 |
| 95 | 49.3 | 89.5 |

Further evaluation of the performance under class imbalance required the calculation of additional metrics that are sensitive to imbalance for the binary classification of Non-IED and IED. The proposed DRU-Mixer reached 89.46% balanced accuracy, 89.17% G-mean, 0.747 MCC, 80.20% F2-score, and 98.03% NPV based on the confusion matrix results. This means the model has maintained a good balance between detecting IED cases and excluding non-IED cases. In addition, the F2-score metric places greater emphasis on recall than on precision, which is important in the clinical setting due to the potential negative consequences of false alarms. The comparison with the cross-entropy and no-weighted-sampler variants further supports the use of focal loss and weighted sampling: both ablation variants yielded higher precision and specificity but lower sensitivity than the full DRU-Mixer configuration. Therefore, the proposed training strategy can be interpreted as a sensitivity-oriented imbalance-aware design rather than a configuration optimized only for overall accuracy.

## 3.2 Multi-Class IED Classification Task

For the multi-class IED classification task, the proposed DRU-Mixer model was evaluated across five IED subtypes: Centro-Parietal IED, Frontal IED, Generalized IED, Occipital IED, and Temporal IED. As shown in Table 3, the model achieved an accuracy of 87.28%, with precision, recall, and F1-score values of 87.92%, 87.78%, and 87.64%, respectively, while maintaining a high specificity of 96.75%.

**Table 3:** Performance evaluation metrics for a multiclass classification task (five IED subtypes)

| | **Accuracy (%)** | **Precision (%)** | **Recall (%)** | **Specificity (%)** | **F1score (%)** |
|---|---|---|---|---|---|
| **DRU-Mixer** | 87.28 | 87.92 | 87.78 | 96.75 | 87.64 |

Fig. 5 illustrates the resulting confusion matrix for the multi-class classification of IED, in which DRU-Mixer was evaluated across five subtypes of IED. The model correctly predicted 65 Centro-Parietal-IED instances, while 8 instances were mistakenly classified as Temporal-IED. 73 samples of Frontal-IED were accurately labeled, although 11 were labeled incorrectly as Generalized-IED and 7 as TemporalIED. For Generalized-IED, 108 samples were correctly classified with very few misclassified as Frontal-IED (5) or Occipital-IED (2). Similarly, Occipital-IED had 79 correct predictions, along with 3 misclassifications of TemporalIED. Temporally, Temporal-IED performed the best with 114 samples being correctly classified, although slight confusions were encountered with Centro-Parietal-IED (8), Frontal-IED (1), Generalized-IED (8), and Occipital-IED (16).

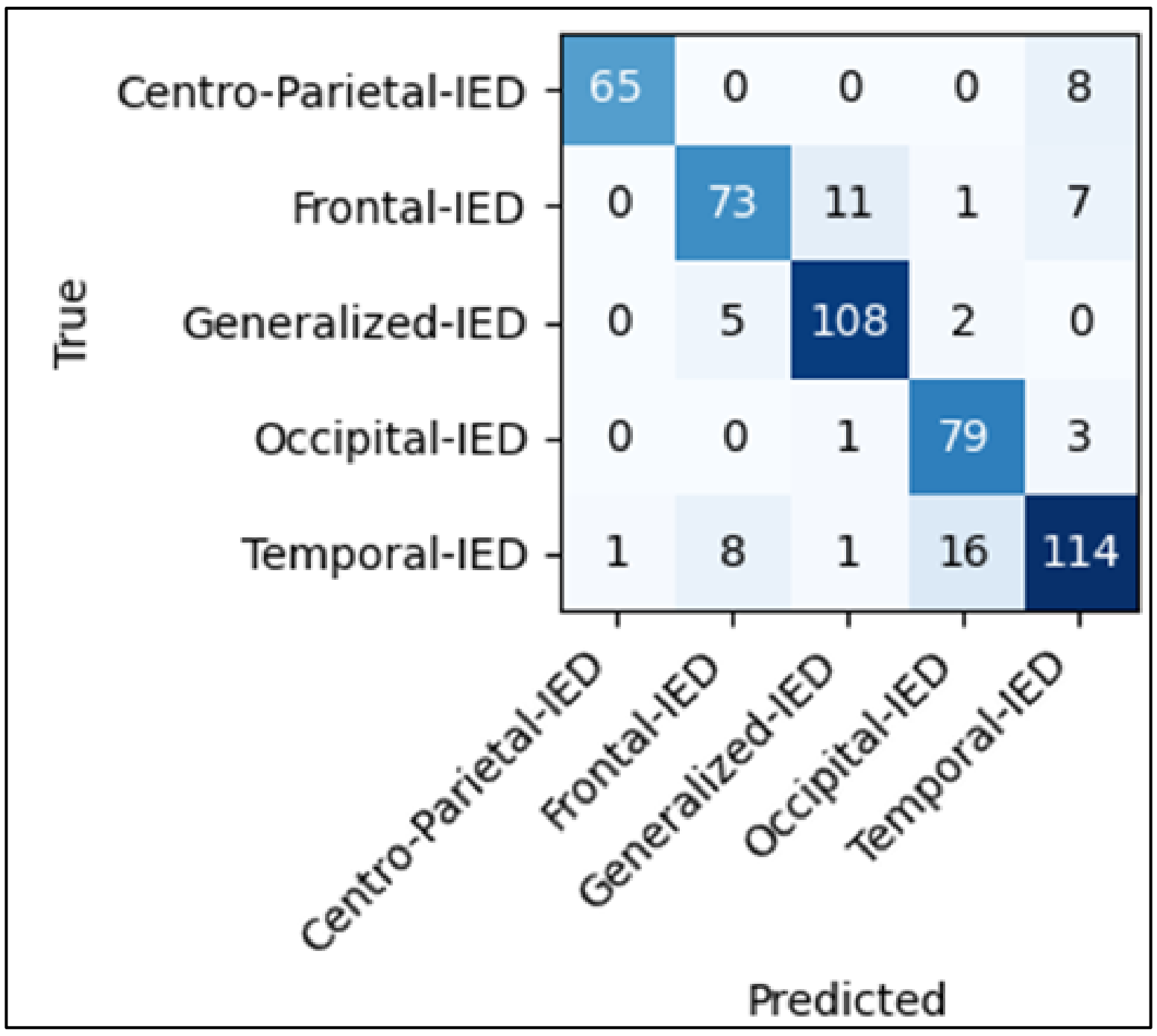

**Figure 5**: Obtained confusion for a multi-class classification task.

The analysis of the performance of the DRU-Mixer classifier on the per-class basis shows that there is some variation in terms of classification accuracy for each of the five types of IED considered. The classifier showed the best results in terms of recall on Occipital-IED (95.18%) and Generalized-IED (93.91%), which demonstrates high sensitivity of classification. In case of Centro-Parietal-IED, recall was equal to 89.04%, while recall for Frontal-IED and Temporal-IED was much lower – 79.35% and 81.43%, respectively, which suggests high frequency of misclassification for these classes. As far as precision is concerned, Centro-Parietal-IED had the best score of 98.48%. Precision for Generalized-IED and Temporal-IED was also strong at 89.26% and 86.36%, whereas Frontal-IED and Occipital-IED recorded slightly lower values of 84.88% and 80.61%.

**Table 4:** Per-class Precision and Sensitivity

| Class | Precision (%) | Sensitivity (%) |
| --- | --- | --- |
| Centro-Parietal-IED | 98.48 | 89.04 |
| Frontal-IED | 84.88 | 79.35 |
| Generalized-IED | 89.26 | 93.91 |
| Occipital-IED | 80.61 | 95.18 |
| Temporal-IED | 86.36 | 81.43 |

## 3.3 Ablation Studies

To evaluate the effect of individual design choices in the DRU-Mixer architecture, we conducted an ablation study with ten experimental variants. The baseline configuration aligns all newly introduced elements, i.e., bandpass filtering, feature normalization, convolutional feature extraction, bidirectional DRU layers, mixer blocks, focal loss, weighted sampling, and Monte Carlo dropout. From this anchor, nine additional variants were defined. First, the mixer block function was examined by excluding them entirely (no mixer) or reducing their quantity to one layer (one mixer). Second, the loss function effect was tested by replacing focal loss with regular cross-entropy (cross-entropy). Third, the sampling strategy was tested by disabling weighted random sampling (no weighted sampler), which allowed us to experiment with the robustness to class imbalance. Fourth, we tested the impact of preprocessing by removing either the bandpass filter (no bandpass) or the standardization step (no Standardization). Fifth, we experimented with sequence modeling choices by constraining the DRU to a unidirectional model (Unidirectional DRU). Sixth, we estimated the impact of convolutional frontend layers by substituting them with a straightforward linear projection (no conv frontend). Seventh and finally, the effect of uncertainty modeling was estimated by disabling Monte Carlo dropout at test time (no MC dropout). All the experiments shared the same training and evaluation pipelines for consistency. Performance was evaluated using accuracy, precision, recall, F1-score, sensitivity, and specificity, thereby enabling a rigorous assessment of how each part of the architecture or methodology contributes to the overall performance of DRU-Mixer in discriminating between IED and non-IED events.

**Table 5**: Results of the ablation study for the DRU-Mixer model.

| **Configuration** | **Accuracy** | **Precision** | **Recall** | **F1-score** | **Sensitivity** | **Specificity** |
|---|---|---|---|---|---|---|
| Baseline | 0.9519 | 0.7276 | 0.8231 | 0.7724 | 0.8231 | 0.9661 |
| No Mixer | 0.9553 | 0.7749 | 0.7734 | 0.7741 | 0.7734 | 0.9753 |
| Cross-Entropy Loss | 0.9577 | 0.8051 | 0.7555 | 0.7795 | 0.7555 | 0.9799 |
| No Weighted Sampler | 0.9596 | 0.8184 | 0.7614 | 0.7889 | 0.7614 | 0.9814 |
| No Bandpass Filter | 0.8436 | 0.2350 | 0.2565 | 0.2452 | 0.2565 | 0.9082 |
| No Standardization | 0.9598 | 0.8108 | 0.7753 | 0.7927 | 0.7753 | 0.9801 |
| Unidirectional DRU | 0.9551 | 0.7907 | 0.7435 | 0.7664 | 0.7435 | 0.9784 |
| No Conv Frontend | 0.9517 | 0.7510 | 0.7674 | 0.7591 | 0.7674 | 0.9720 |
| No MC Dropout | 0.9608 | 0.8276 | 0.7634 | 0.7942 | 0.7634 | 0.9825 |
| One Mixer Layer | 0.9600 | 0.7918 | 0.8091 | 0.8004 | 0.8091 | 0.9766 |

Table 5 provides the ablation study outcome of the DRUMixer model under ten varied settings. The baseline setting achieved balanced performance with an accuracy of 0.9519, a recall of 0.8231, and an F1-score of 0.7724, serving as a reference. Once the mixer layers were removed (no mixer), precision slightly improved to 0.9553 and accuracy reached 0.7749, though recall was lowered to 0.7734, showing that mixers gain additional recall. Cross-entropy loss instead of focal loss increased precision to 0.8051 and specificity to 0.9799, while recall fell to 0.7555, substantiating the advantage of using focal loss in addressing class imbalance. Eliminating weighted sampling (no weighted sampler) yielded the highest precision of 0.8184 and the highest specificity of 0.9814; however, recall was moderately reduced to 0.7614, indicating that the sampler helps in returning more positive cases at the expense of precision. The removal of band pass filtering (or lack of band pass processing) caused the biggest decrease in the performance, where the accuracy was decreased to 0.8436, precision to 0.2350, and the F1-score to 0.2452, thus, showing the key importance of the frequency-domain pre-processing. The exclusion of the standardization process (lack of standardization) resulted in close-to-baseline results, where the accuracy was equal to 0.9598, and the precision and recall rates were balanced, thus, proving the secondary importance of the standardization process. The use of the DRU instead of GRU resulted in decreased recall rate of 0.7435 and F1-score of 0.7664, which shows the importance of bidirectionality for modeling temporal contexts. Similarly, switching from the convolutional frontend to the light projection frontend (No conv frontend) led to decreased precision of 0.7510 and F1-score of 0.7591, which proves the importance of the convolutional frontend. Using Monte Carlo dropout resulted in the best precision rate of 0.8276 and best accuracy rate of 0.9608 but reduced the recall to 0.7634, thus, proving that uncertainty modeling is not universally beneficial for balanced datasets. Finally, using a single mixer layer resulted in accuracy of 0.9600 and recall of 0.8091 with the highest F1-score of 0.8004.

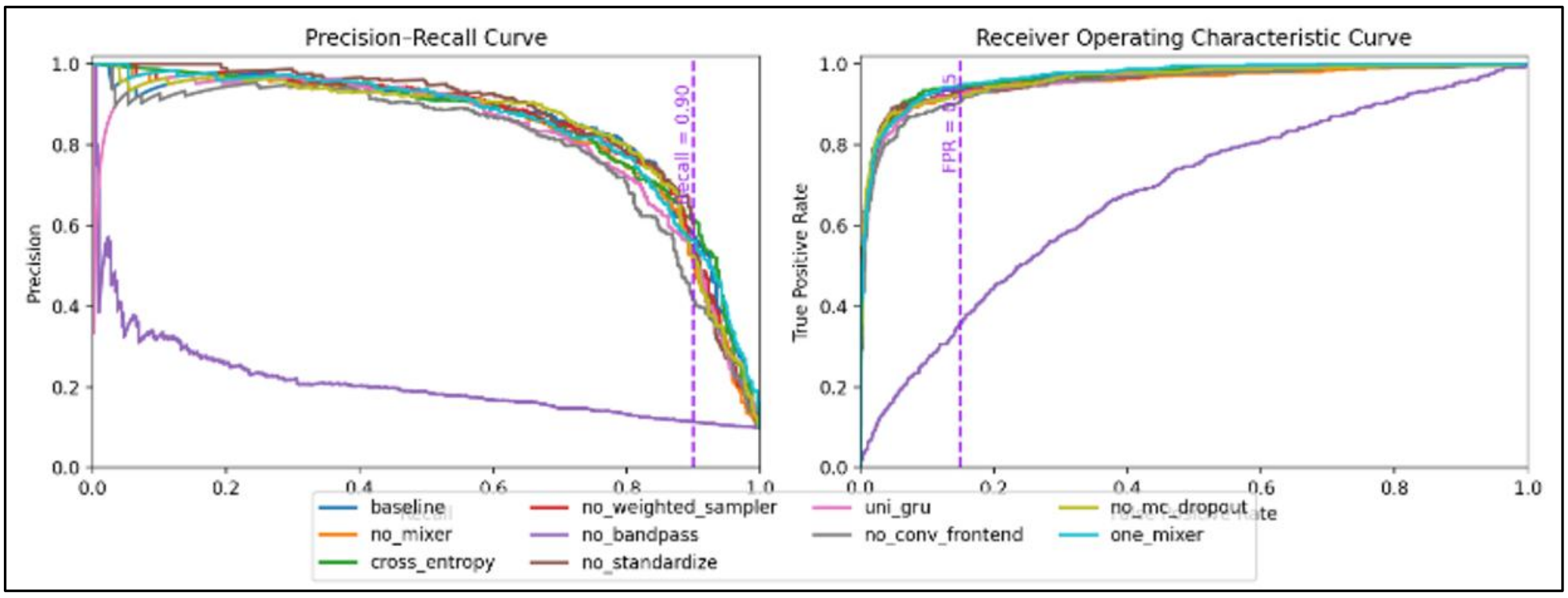

**Figure 6**: Validation results of different model configurations from the ablation study.

The results of the validation procedure of the ablation study are shown in Figure 6 using Precision-Recall (PR) and Receiver Operating Characteristic (ROC) analysis for the different DRU-Mixer implementations. The baseline model has reached an equilibrium of precision and recall, and this trend is observed by almost all other configurations, among them: no mixer, cross-entropy loss, no weighted sampler, no standardization, one-directional DRU, no convolutional front-end, no MC dropout, and a one-layered mixer. In contrast to that, the omission of the bandpass filter causes a severe degradation in both PR and ROC curves, which shows the importance of the step in the preprocessing stage. It is worthy of note that configurations like no weighted sampler and no MC dropout have similar or even slightly increased precision at the high values of recall, which proves their stability across a wider range of thresholds. These claims are supported by ROC analysis, where almost all models have close areas under the curve (AUC) compared to the baseline except for no bandpass, which has significantly lower TPR at all FPR.

It should be noted that the full DRU-Mixer configuration did not achieve the highest value for every individual metric. Some ablation variants, such as No MC Dropout and One Mixer Layer, produced higher accuracy, precision, specificity, or F1-score values. However, these improvements were accompanied by a reduction in sensitivity compared with the full

DRU-Mixer model. Since IED detection is clinically sensitive to missed epileptiform events, recall/sensitivity is an important evaluation criterion. The full DRU-Mixer achieved the highest sensitivity among the tested configurations while maintaining competitive accuracy, specificity, and F1-score. Therefore, the contribution of the full architecture should be interpreted as providing a sensitivity-oriented and clinically conservative configuration rather than maximizing all metrics simultaneously.

## 3.4 Calibration and Uncertainty Analysis

To quantitatively support the uncertainty-aware inference claim, calibration and uncertainty analyses were conducted using deterministic and MC-dropout-based inference. In MC-dropout inference, multiple stochastic forward passes were performed at test time, and the mean class probability was used for final prediction. The dispersion of stochastic predictions was further used to estimate predictive uncertainty. Calibration was evaluated using expected calibration error (ECE), maximum calibration error (MCE), Brier score, and negative log-likelihood (NLL). In addition, predictive entropy, positive-class probability variance, and mutual information were computed to quantify predictive uncertainty. Finally, uncertainty values were compared between correctly and incorrectly classified samples, and error-detection AUROC was calculated to assess whether uncertainty measures could identify unreliable predictions.

**Table 6.** Calibration and uncertainty-aware inference performance of deterministic and MC-dropout-based DRU-Mixer.

| Inference mode | ROC-AUC | PR-AUC | Brier | NLL | ECE | MCE |
|---|---|---|---|---|---|---|
| Deterministic | 0.9558 | 0.8253 | 0.0398 | 0.2652 | 0.0392 | 0.7491 |
| MC dropout | 0.9643 | 0.8415 | 0.0364 | 0.2377 | 0.0334 | 0.2699 |

Table 6 presents the calibration and probability-quality performance of deterministic and MC-dropout-based DRU-Mixer inference. Compared with deterministic inference, MC dropout improved both threshold-independent discrimination metrics, increasing ROC-AUC from 0.9558 to 0.9643 and PR-AUC from 0.8253 to 0.8415. More importantly, all probability-based calibration metrics improved consistently. The Brier score decreased from 0.0398 to 0.0364, NLL decreased from 0.2652 to 0.2377, ECE decreased from 0.0392 to 0.0334, and MCE decreased from 0.7491 to 0.2699. These results indicate that MC dropout produced better calibrated probability estimates and more reliable confidence scores. Therefore, MC dropout should be interpreted primarily as a confidence-aware inference mechanism rather than as a strategy aimed only at improving conventional classification metrics.

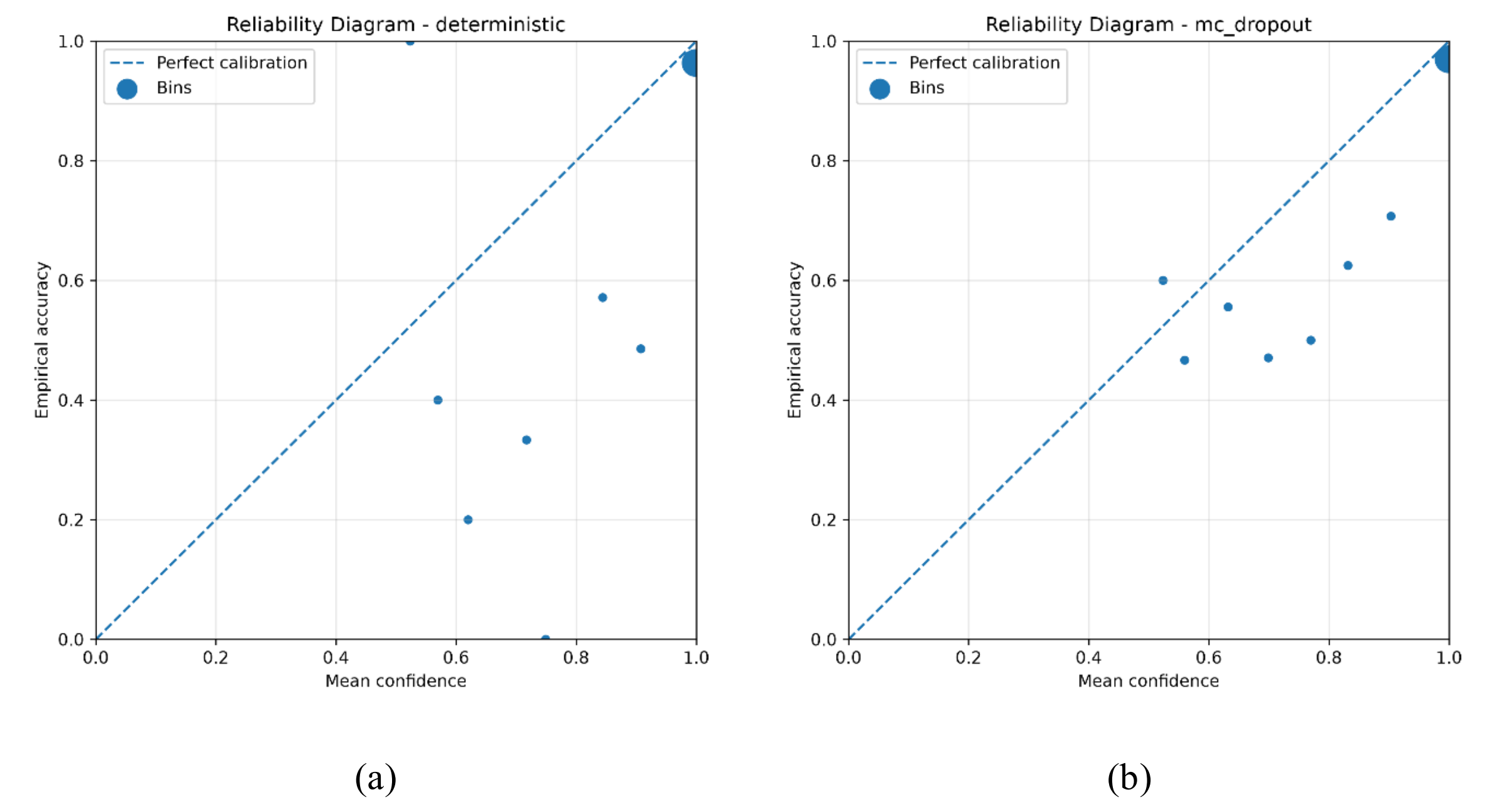

**Figure 7.** Reliability diagrams for deterministic and MC-dropout-based DRU-Mixer inference.

Figure 7 shows the reliability diagrams of deterministic and MC-dropout-based inference. The MC dropout curve is closer to the ideal calibration line, indicating better agreement between predicted confidence and empirical accuracy. This visual result is consistent with the lower ECE, MCE, Brier score, and NLL values obtained by MC dropout.

**Table 7.** Uncertainty comparison between correctly and incorrectly classified test samples under MC dropout.

| Uncertainty measure | Correct predictions | Incorrect predictions | Error-detection AUROC |
|---|---|---|---|
| Predictive entropy | 0.0120 ± 0.0666 | **0.1690 ± 0.2364** | 0.8910 |
| Positive probability variance | 0.001513 ± 0.012918 | **0.028303 ± 0.050701** | 0.8939 |
| Mutual information | 0.004409 ± 0.034171 | **0.076606 ± 0.128339** | 0.8947 |

Table 7 compares the uncertainty values of correctly and incorrectly classified test samples under MC dropout. Incorrect predictions showed substantially higher predictive entropy, positive-class probability variance, and mutual information than correct predictions. Predictive entropy increased from 0.0120 ± 0.0666 in correct predictions to 0.1690 ± 0.2364 in incorrect predictions. Similarly, MC probability variance increased from 0.001513 ± 0.012918 to 0.028303 ± 0.050701, and mutual information increased from 0.004409 ± 0.034171 to 0.076606 ± 0.128339. The uncertainty-based error-detection AUROC values were approximately 0.89 for all three uncertainty measures.

**Table 8.** Quantitative occlusion-faithfulness analysis of DRU-Mixer.

| Occlusion setting | n | Mean base P(IED) | Mean occluded P(IED) | Mean ΔP(IED) | Relative drop (%) |
|---|---|---|---|---|---|
| Top 5% most important regions | 10 | 0.9938 | 0.0859 | 0.9078 ± 0.1701 | 91.34 ± 17.11 |
| Top 10% most important regions | 10 | 0.9938 | 0.1207 | 0.8731 ± 0.2857 | 87.83 ± 28.76 |
| Random 10% regions | 10 | 0.9938 | 0.7155 | 0.2783 ± 0.2321 | 28.13 ± 23.57 |
| Least important 10% regions | 10 | 0.9938 | 0.9982 | -0.0044 ± 0.0060 | -0.45 ± 0.62 |

Table 8 reports the quantitative occlusion-faithfulness results obtained from correctly classified IED samples. Masking the top 5% of channel-time regions with the highest occlusion importance caused the mean predicted IED probability to decrease from 0.9938 to 0.0859, corresponding to a 91.34% relative reduction. A similar pattern was observed when the top 10% of important regions were masked, where the predicted IED probability decreased by 87.83%. By contrast, masking randomly selected 10% regions resulted in a much smaller decrease of 28.13%, and masking the least important 10% regions had almost no effect on the predicted IED probability.

## 3.5 Explainability Analysis

The explainability analysis is based on occlusion mapping, which involves a structured and measured evaluation of the influence of each time interval and channel on the model's IED probability [27]. In one-dimensional attribution, the input sequence is scanned by a window, and the signal in each window is replaced with zeros (or the local mean). The difference in IED probability predicted before and after occlusion is captured and interpolated to obtain a dense curve of importance in time. We extend this to a channel × time map by occluding each channel one at a time at every window position, aggregating the drop in probability and normalizing it over the number of hits [28]. The resulting heatmaps are visualized over the raw, unprocessed, bandpass-filtered, and normalized traces of the 23 scalp channels, allowing for identification of the most significant temporal windows and spatial electrodes that strongly influence positive decisions. This structure does not modify the model and training pipeline, preserves attribution in the task units (ΔP for IED), and is class-imbalance stable, as it perturbs the inputs and not the internal parameters. Uncertainty-aware inference also improves interpretability. At test time, Monte Carlo dropout takes N stochastic forward passes and averages them to obtain class probabilities and a base confidence associated with each explanation. High ΔP regions on occlusion maps, therefore, indicate features that robustly raise IED probability under distributional noise, whereas diffuse or weak attributions indicate brittle evidence.

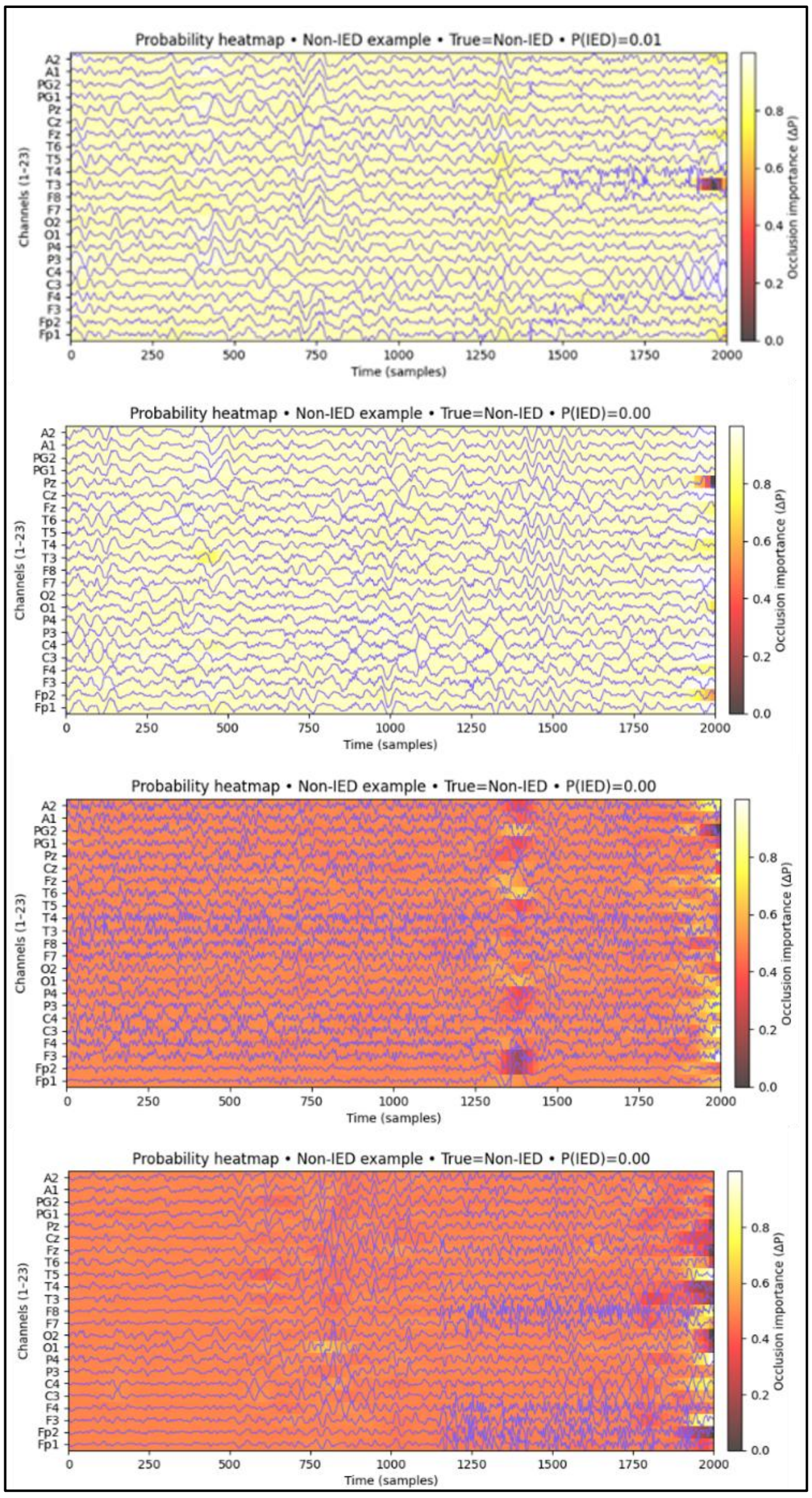

**Figure 8**: Probability heatmap obtained with the occlusion visualization technique on Non -IED samples.

The occlusion maps for the non-IED and IED cases show clear differences in how the DRU-Mixer model allocates importance across time and channels. As seen in Fig. 8, in the Non-IED cases, the probability heatmaps show sparse and low activation overall, with only localized and scattered importance signals. The ΔP values are weak, dispersed, and lack consistency across channels, often manifesting as transient spikes in posterior leads, such as Pz, T6, or PG2. However, these areas are not maintained and do not spread through several channels at once, which demonstrates the certainty of the model in classifying these areas as NonIED with a near-zero value for P(IED). It is seen from this that in the presence of no epileptic activity, the discriminative importance of variations in EEG becomes low for the model.

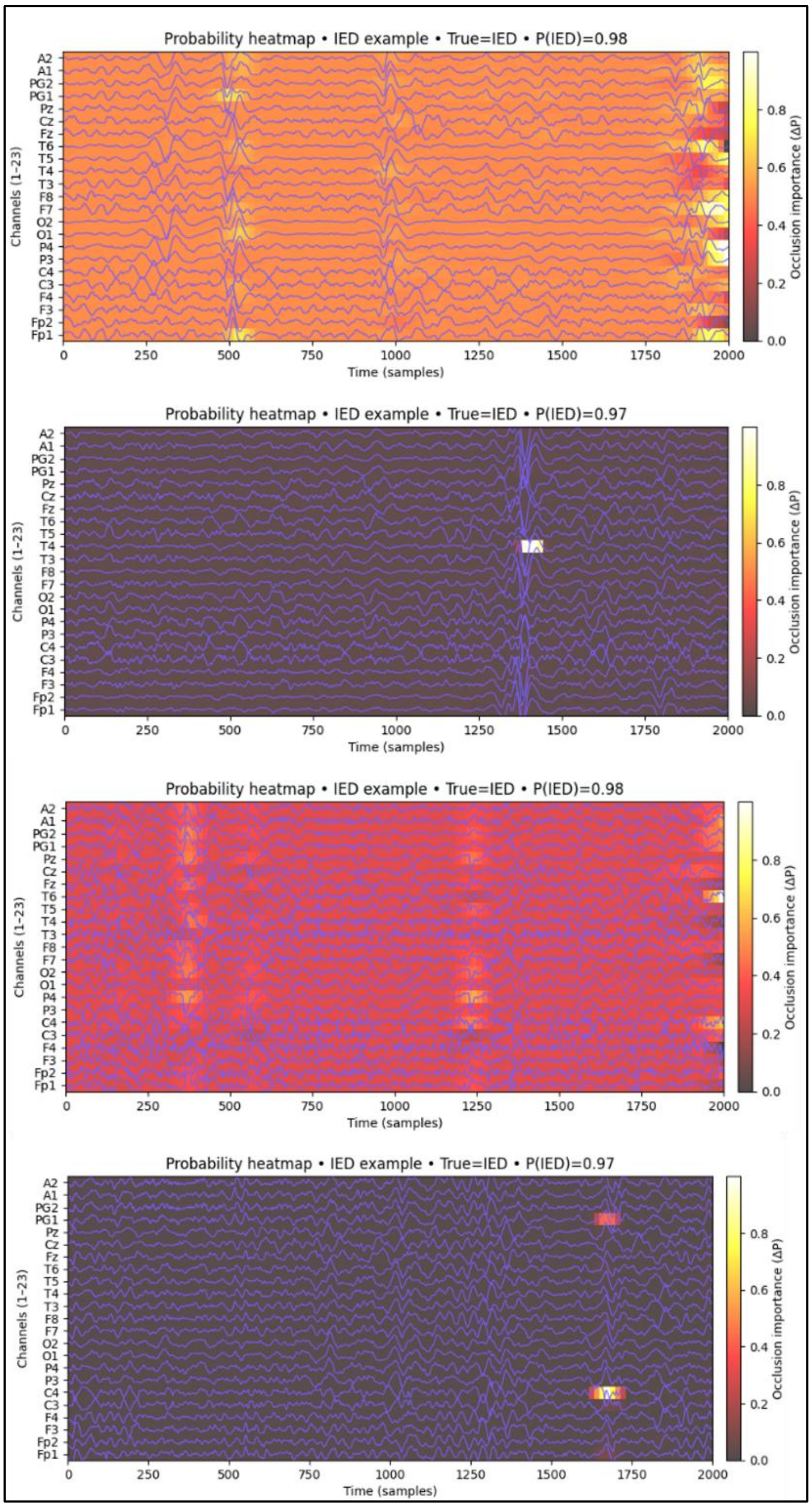

**Figure 9**: Probability heatmap obtained with the occlusion visualization technique on IED samples.

In contrast, as given in Fig. 9, IED occlusion maps reveal strong, spatially and temporally coherent clusters of importance. Characteristic high ΔP values occur in multiple adjacent channels, particularly temporal and central regions, following the morphological fields of IED discharges. These activations have extended time windows, and their removal decreases the estimated IED probability substantially (P(IED) ≈ 0.97–0.98). Unlike the dispersed signals found in Non-IED, the IED maps contain concentrated regions that match the spike-and-wave structure of the IED, and thus the clinical information has been identified by the model. These comparisons show that DRU-Mixer is successful in differentiating between Non-IED and IED because of focusing on structured, multi-channel temporal activity of the epileptiforms.

# 4. Discussion

The current paper proposes a novel deep learning algorithm called DRU-Mixer for automatic identification and classification of IED from multi-channel electroencephalography (EEG) data. DRU-Mixer is built based on the integration of convolutional layers for extraction of local features, a dilated recurrent unit for capturing temporal dependencies, and sequential mixer blocks for increasing interactions of features at both token and channel levels. Bandpass filter and channel standardization were thoroughly performed during the preprocessing stage. Focal loss was implemented to deal with class imbalance. To increase the interpretability of DRU-Mixer, the occlusion-based XAI approach was used for systematic evaluation of the importance of different EEG channels and temporal windows for the performance of the model. Through masking specific parts of the input sequence and observing the changes in the model predictions, this method allows obtaining easy-to-understand information about the most informative regions of EEG recordings for the identification of IED. This type of interpretability ensures that the model makes its decisions based on physiologically meaningful information instead of irrelevant artifacts and gives guidance on the spatial and temporal importance of identified abnormalities for clinicians. Moreover, introducing an occlusion-based XAI approach to the diagnosis process helps to increase transparency and trustworthiness in the process of diagnosing epilepsy.

For binary classification between IED and non-IED segments, DRU-Mixer achieved 95.19% accuracy, with 72.76% precision, 82.31% recall, 96.61% specificity, and 77.24% F1-score. Sensitivity-targeted analysis also demonstrated consistent precision-specificity trade-offs across a range of clinically relevant thresholds. When addressing the multiclass situation involving five subtypes of IED (centro-parietal, frontal, generalized, occipital, and temporal IED), the model maintained robust performance with an accuracy of 87.28%, a precision of 87.92%, a recall of 87.78%, a specificity of 96.75%, and an F1-score of 87.64%. Confusion matrix analysis confirmed strong discrimination across all classes with excellent recall for generalized and temporal IED.

Table 8 summarizes the performance of the suggested DRU-Mixer model compared to state-of-the-art VGGEEG, VGGSleep, and vEpiNet methods on the same dataset, for sensitivity rates ranging from 50% to 95%. The results indicate that DRU-Mixer achieves precision and specificity comparable to or better than existing models, with a particularly noticeable advantage in the middle sensitivity level (65–80%), where it maintains precision above 76.9% and specificity above 97.4%. While VGGEEG and VGGSleep are slightly more accurate at lower operating thresholds of sensitivity (50–60%), their accuracy declines more drastically as sensitivity increases. In the meantime, vEpiNet performs worst in precision at the majority of operating points of sensitivity, with a precision as low as 24.5% and 95% sensitivity, showing its instability at strict operating points.

For transparency, the comparison in Table 9 should be interpreted as a sensitivity-matched benchmark comparison on the same IED dataset. DRU-Mixer was evaluated using the segment-level stratified 80/20 train-test split, 2000-sample EEG windows, band-pass filtering, channel-wise standardization, and threshold sweeping across predefined sensitivity levels. The comparison values for VGGEEG, VGGSleep, and vEpiNet were taken from the benchmark results reported in [15] at matched sensitivity operating points. Therefore, although all methods are compared on the same dataset and at the same sensitivity levels, the comparison should not be interpreted as a fully controlled head-to-head reimplementation unless identical train-test splits, preprocessing, windowing, and threshold-selection protocols are explicitly available for all methods.

**Table 9**. Sensitivity-matched benchmark comparison of DRU-Mixer with existing methods on the same IED dataset.

| Sensitivity | VGGEEG Precision (%) [15] | VGGEEG Specificity (%) [15] | VGGSleep Precision (%) [15] | VGGSleep Specificity (%) [15] | vEpiNet Precision (%) [15] | vEpiNet Specificity (%) [15] | **DRU-Mixer Precision (%)** | **DRU-Mixer Specificity (%)** |
|---|---|---|---|---|---|---|---|---|
| 50% | 96.1 | 99.8 | 96.5 | 99.8 | 93.9 | 99.6 | **93.9** | **99.7** |
| 55% | 94.7 | 99.7 | 95.1 | 99.7 | 91.9 | 99.5 | **91.7** | **99.5** |
| 60% | 94.0 | 99.6 | 91.6 | 99.4 | 88.8 | 99.2 | **89.3** | **99.2** |
| 65% | 91.9 | 99.4 | 90.4 | 99.3 | 84.7 | 98.7 | **87.0** | **98.9** |
| 70% | 87.8 | 99.0 | 86.2 | 98.8 | 79.0 | 98.0 | **83.8** | **98.5** |
| 75% | 82.3 | 98.3 | 83.1 | 98.4 | 73.1 | 97.0 | **81.7** | **98.2** |
| 80% | 74.2 | 97.0 | 76.3 | 97.3 | 63.6 | 95.0 | **76.9** | **97.4** |
| 85% | 63.4 | 94.8 | 67.1 | 95.5 | 53.8 | 92.0 | **66.6** | **95.3** |
| 90% | 50.4 | 90.5 | 60.0 | 93.6 | 41.1 | 85.9 | **54.8** | **91.8** |
| 95% | 29.5 | 75.8 | 49.4 | 89.6 | 24.5 | 67.9 | **49.3** | **89.5** |

The curves in Fig. 10 highlight the relative performance of DRU-Mixer compared to VGGEEG, VGGSleep, and vEpiNet across different levels of sensitivity. The precision–sensitivity plot shows that DRU-Mixer maintains high precision values throughout and remains competitive with VGGEEG and VGGSleep, but is superior to vEpiNet, especially at higher sensitivity thresholds. Similarly, under the specificity–sensitivity plot, DRU-Mixer remains highly robust, with near-optimal specificity consistent with VGGEEG and VGGSleep and superior to vEpiNet at all levels of sensitivity.

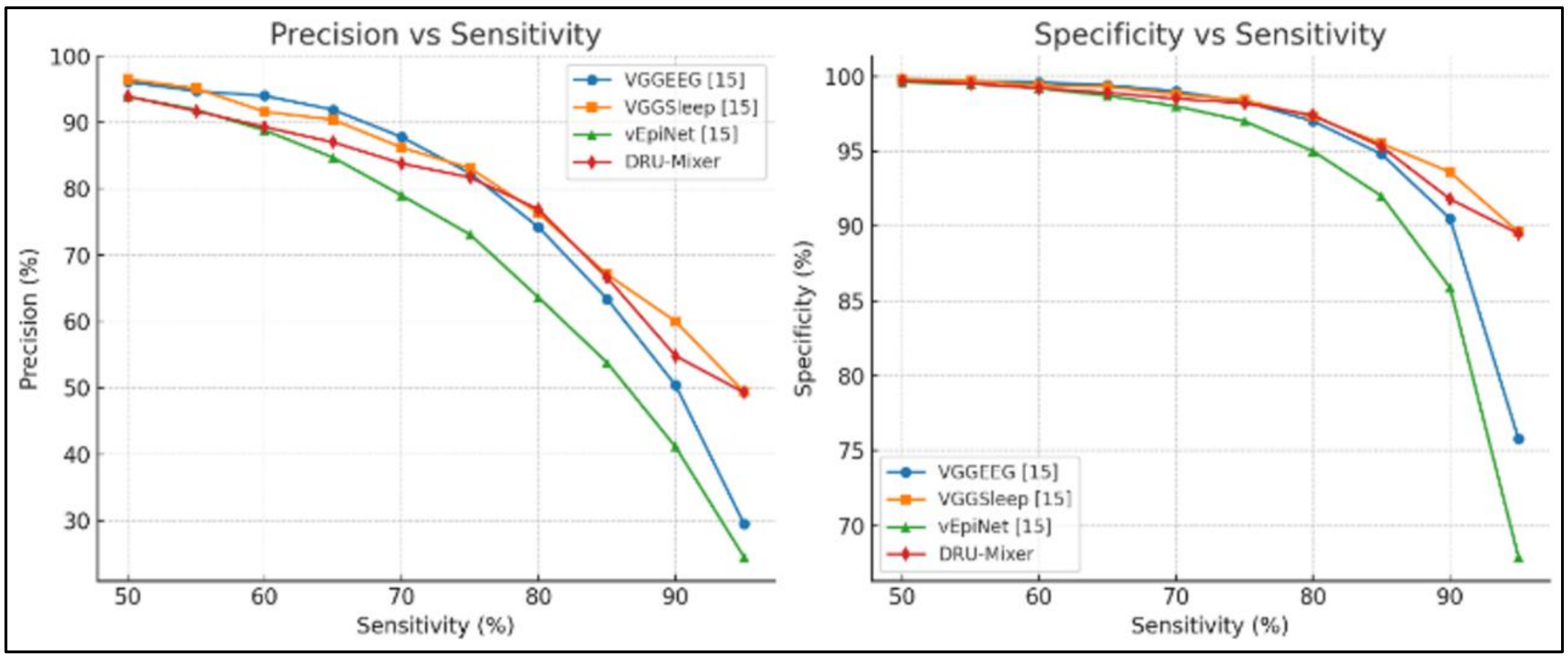

**Figure 10**: Precision and specificity across sensitivity levels for DRU-Mixer compared with VGGEEG [15], VGGSleep [15], and vEpiNet [15].

The main advantages of the suggested DRU-Mixer method are:

1) There are consistent high precision and specificity across different sensitivity levels, and the model provides good generalization for binary as well as multiclass classification tasks.

2) Through convolutional layers, DRU, and mixer blocks, the model can learn spatial (channel-wise) and temporal relationships in EEG signals.

3) Occlusion explainability provides clear intuitions on what channels and slots have the most significant influence in decisions, enabling trust and clinical adoption.

4) Modularity in architecture allows ablation and parameter tuning modifications (e.g., with/without mixer, changing loss functions), enabling the method to be versatile across varying experiment settings.

In addition, the constraints of the DRU-Mixer are:

1) There is added training and inference time through the incorporation of convolution, DRU, and mixer layers compared to more lightweight baselines.

2) High-quality, preprocessed EEG data are very much dependent on high performance and unbalanced or damaged data can lower reliability.

3) Tuning of model dropout, dilation, number of mixer layers, and loss function choices must be made, which could render it difficult to deploy.

4) The train-test split was performed at the EEG segment level rather than at the patient level. Although this setting evaluates segment-level discriminative performance, future studies should further validate DRU-Mixer using patient-level or cross-patient evaluation protocols to assess its generalization ability to unseen patients.

While it performs well on benchmark data, optimization for real-time clinical application with noisy and heterogeneous EEG recordings would be needed.

The uncertainty analysis further clarifies the role of MC dropout in the proposed framework. Although MC dropout should not be interpreted as a mechanism that uniformly improves all conventional classification metrics, it provided better calibrated probability estimates and meaningful uncertainty information. Compared with deterministic inference, MC dropout reduced ECE, MCE, Brier score, and NLL, suggesting improved confidence calibration. In addition, incorrectly classified samples exhibited substantially higher entropy, probability variance, and mutual information than correctly classified samples. This finding is clinically relevant because uncertain or unreliable predictions can be flagged for additional expert review rather than being treated as fully confident automated decisions.

## 5. Conclusion

The DRU-Mixer model introduced in this work demonstrates strong performance on both binary and multiclass classification tasks by combining convolutional, recurrent, and mixer components within a single framework. Through thorough ablation studies and comparisons to the state-of-the-art, the model demonstrated effectiveness in learning spatiotemporal patterns in EEG signals, while also providing explainability through occlusion-based analysis. The results confirm that the proposed design is balanced in precision, sensitivity, and specificity and show an improvement over several well-known baselines. Future research should be tested on larger and more diverse datasets to determine its robustness in various recording situations and patient populations. Exploring lighter or optimized versions of the model would be worthwhile for real-time clinical deployment, particularly in resource-constrained environments. Some additional directions include leveraging multimodal cues such as video or other biosignals, enhancing methods for explainability beyond occlusion maps, and investigating self-supervised or transfer learning mechanisms to reduce reliance on annotated data. These actions would further the applicability of the DRU-Mixer in real-world clinical and research settings.

## Acknowledgement

The first author wishes to acknowledge the project funding from the University of Southern Queensland through providing an International PhD Stipend Research Scholarship and University of Southern Queensland International Fee Research Scholarship (2023-2026).

## Fundings

## Ethics Declaration

Not applicable.